\documentclass[intlimits,twoside,a4paper]{article}

\usepackage{subfig}
\usepackage[T2A]{fontenc}
\usepackage[cp1251]{inputenc}

\usepackage[eqsecnum]{cmpj3}

\issue{2021}{24}{2}{23703}
\doinumber{10.5488/CMP.24.23703}

\title[First-principles investigation of half-metallic ferromagnetism]%
{First-principles investigation of half-metallic ferromagnetism of Fe$_2$YSn (Y=Mn, Ti and V) Heusler alloys}%

\author[M. Sayah \textsl{et al.}]{M. Sayah\refaddr{label1}, S. Zeffane \refaddr{label1}, M. Mokhtari\refaddr{label1,label2}, F. Dahmane\refaddr{label1,label3}, L. Zekri\refaddr{label2}, R. Khenata\refaddr{label3}, N.~Zekri\refaddr{label2}}
\addresses{
\addr{label1} D\'epartement de SM, Institut des Sciences et des Technologies, Centre Universitaire de Tissemsilt, 38000 Tissemsilt, Alg\'erie 
\addr{label2} Universit\'e des Sciences et de la Technologie d'Oran Mohamed Boudiaf, USTO-MB, LEPM, BP 1505, El M' Naouar, 31000 Oran, Algeria
\addr{label3} Laboratoire de Physique Quantique et de Mod\'elisation Math\'ematique (LPQ3M), D\'epartement de Technologie, Universit\'e de Mascara, 29000 Mascara, Alg\'erie  }

\date{Received April 28, 2020, in final form February 22, 2021}
\begin{document}

\maketitle
\begin{abstract}

In this paper, we use the first-principles calculations based on the density functional theory to investigate structural, electronic and magnetic properties of Fe$_{2}$YSn with (Y~=~Mn, Ti and V). The generalized gradient approximation (GGA) method is used for calculations. The Cu$_{2}$MnAl type structure is energetically more stable than the Hg$_{2}$CuTi type structure. The negative formation energy is shown as the evidence of thermodynamic stability of the alloy. The calculated total spin moment is found as 3$\mu_\text{B}$ and 0$\mu_\text{B}$  at the equilibrium lattice constant for Fe$_{2}$MnSn and Fe$_{2}$TiSn respectively, which agrees with the Slater-Pauling rule of $M_t= Z_t-24$. The study of electronic and magnetic properties proves that Fe$_{2}$MnSn and Fe$_{2}$TiSn full-Heusler alloys are complete half-metallic ferromagnetic materials
\end{abstract}

\section{Introduction}

In 1983, de Groot et al~\cite{1}  established the half-metallicity in NiMnSb half Heusler alloy. Later, through theoretical calculations and experiments, many compounds were found to be half-metals, including Heusler alloys~\cite{2,3}, alkali metal or transition metal chalcogenides~\cite{4}, doped diluted magnetic semiconductors~\cite{5,6}, zinc-blende and wurtzite structural compounds~\cite{7}. 
Considering the standing great potential advantages, spintronics still faces some challenges, such as generation of high spin injectors~\cite{8}. 
%
%
In recent years, Heusler alloys received an insistent attention due to their interesting physical properties~\cite{9,10,11}, their remarkable electronic structure makes it possible to use them in various spintronic devices such as spin-transfer torque and large magneto-resistance spinvalves devices~\cite{12}.
Heusler alloys are ternary inter-metallic compounds, which were first discovered by Heusler in 1903~\cite{13}. This remarkable material and its relatives, which by now comprise a vast collection of more than 1000 compounds, are now known as Heusler compounds. They are ternary semiconducting or metallic materials with a 1:1:1 (also known as ``half-Heusler'') or a 2:1:1 stoichiometry (also known as ``full-Heusler'')~\cite{13}. Several Fe-based Heusler alloys have already been studied, though due to the differences in their experimental and theoretical results further investigations are still being carried out~\cite{13,3}.
This paper is structured as follows: in section~\ref{s2}, we briefly describe the computational method used in this work. Results and discussions of our study are presented in section~\ref{s3}. Finally, a summary of the work is given in section~\ref{s4}.

\section{Method of description}\label{s2}
The first-principle calculations of Fe$_{2}$YSn (Y~=~Mn, Ti, and V) alloys are performed based on the density functional theory (DFT)~\cite{14}, which is implemented in WIEN2k code~\cite{15}. The solution of the Kohn-Sham equation~\cite{14} is  done using the full potential linearized augmented plane wave (FP-LAPW) method~\cite{15}. The exchange correlation potential is calculated using the Perdew-Burke-Ernzerhof parameterization of the generalized gradient approximation PBE-GGA~\cite{16}. In the calculations reported in this paper, we use a parameter $RMT\times K_{\text{max}} =8$, which defines the matrix size convergence, where $K_{\text{max}}$ is the plane wave cut-off and $RMT$ is the smallest of all atomic sphere radii. In the full potential scheme, the whole crystal is divided into two different parts: the first part is the atomic sphere while the second part includes the interstitial regions. Moreover, the valence wave function inside the muffin-tin (MT) sphere was expanded up to $I_{\text{max}} =10$, while the charge density was Fourier expanded up to $G_{\text{max}} =12$~a.u$^{-1}$. The self-consistent calculations are considered to be converged when the total energy of the system is stable within $10^{-4}$~Ry.

\section{Results and discussions}\label{s3}

\begin{figure}%
    \centering
    \subfloat[]{{\includegraphics[width=8cm]{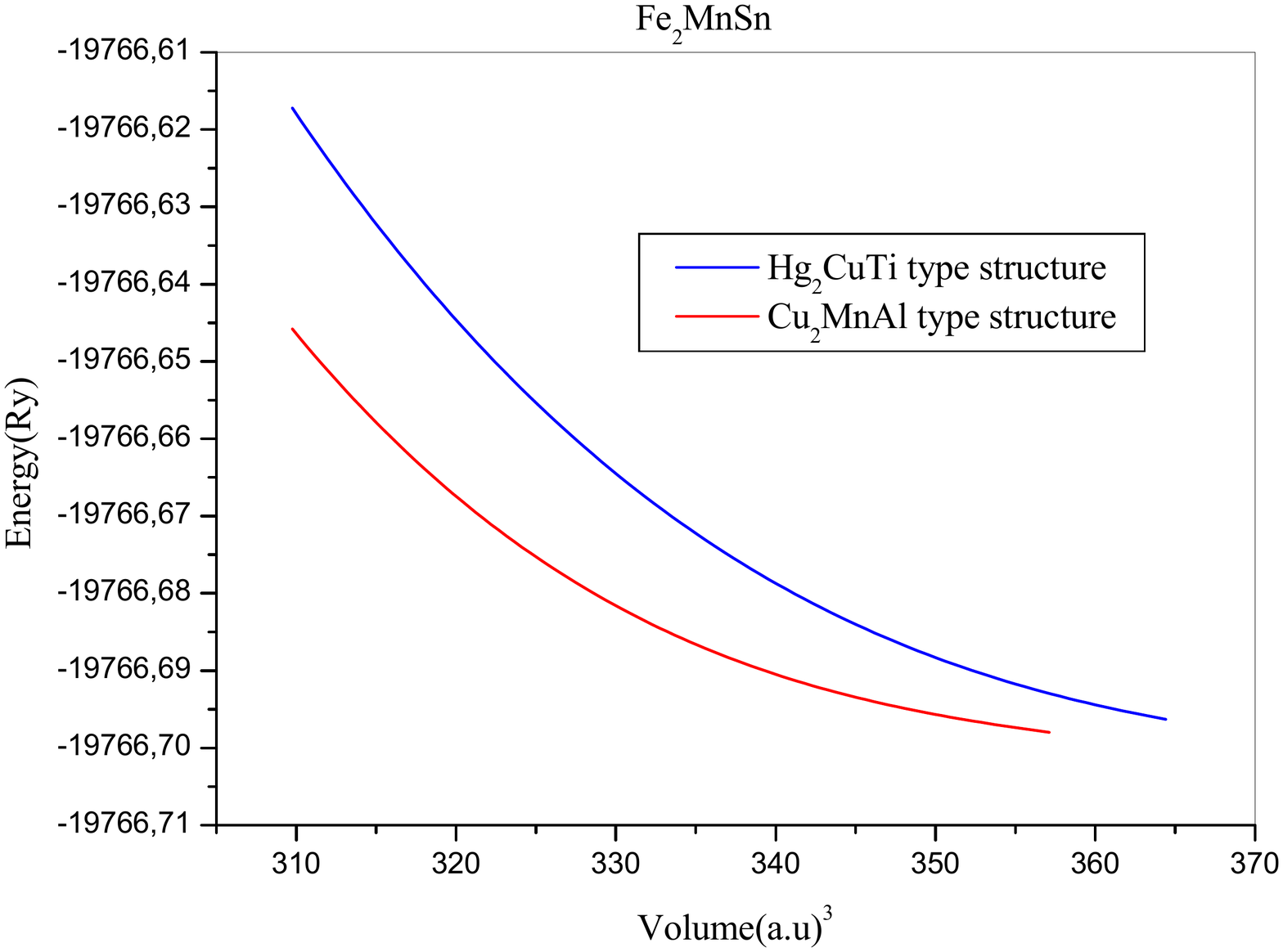} }}%
    \qquad
    \subfloat[]{{\includegraphics[width=8cm]{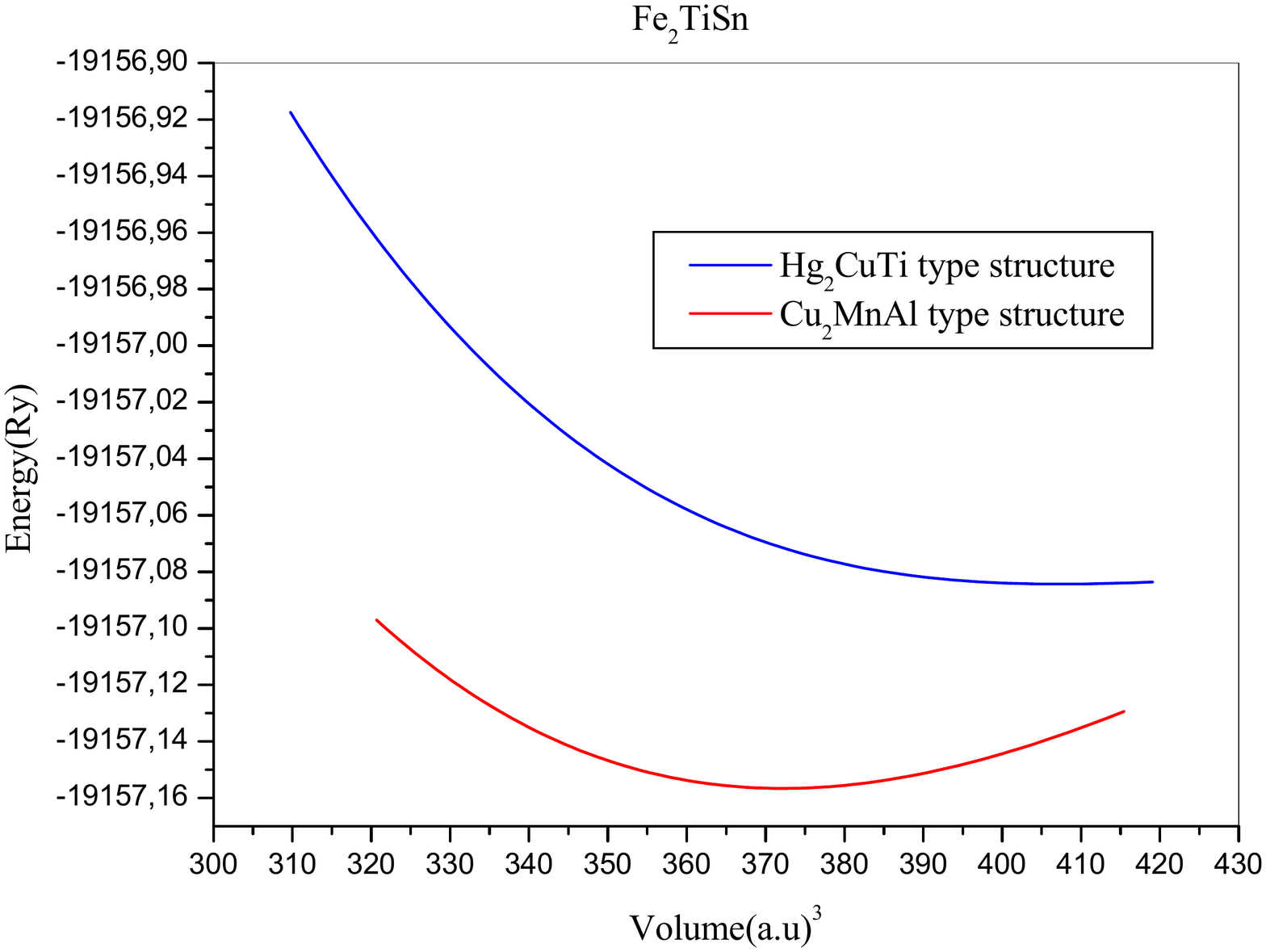} }}%
 \qquad
    \subfloat[]{{\includegraphics[width=8cm]{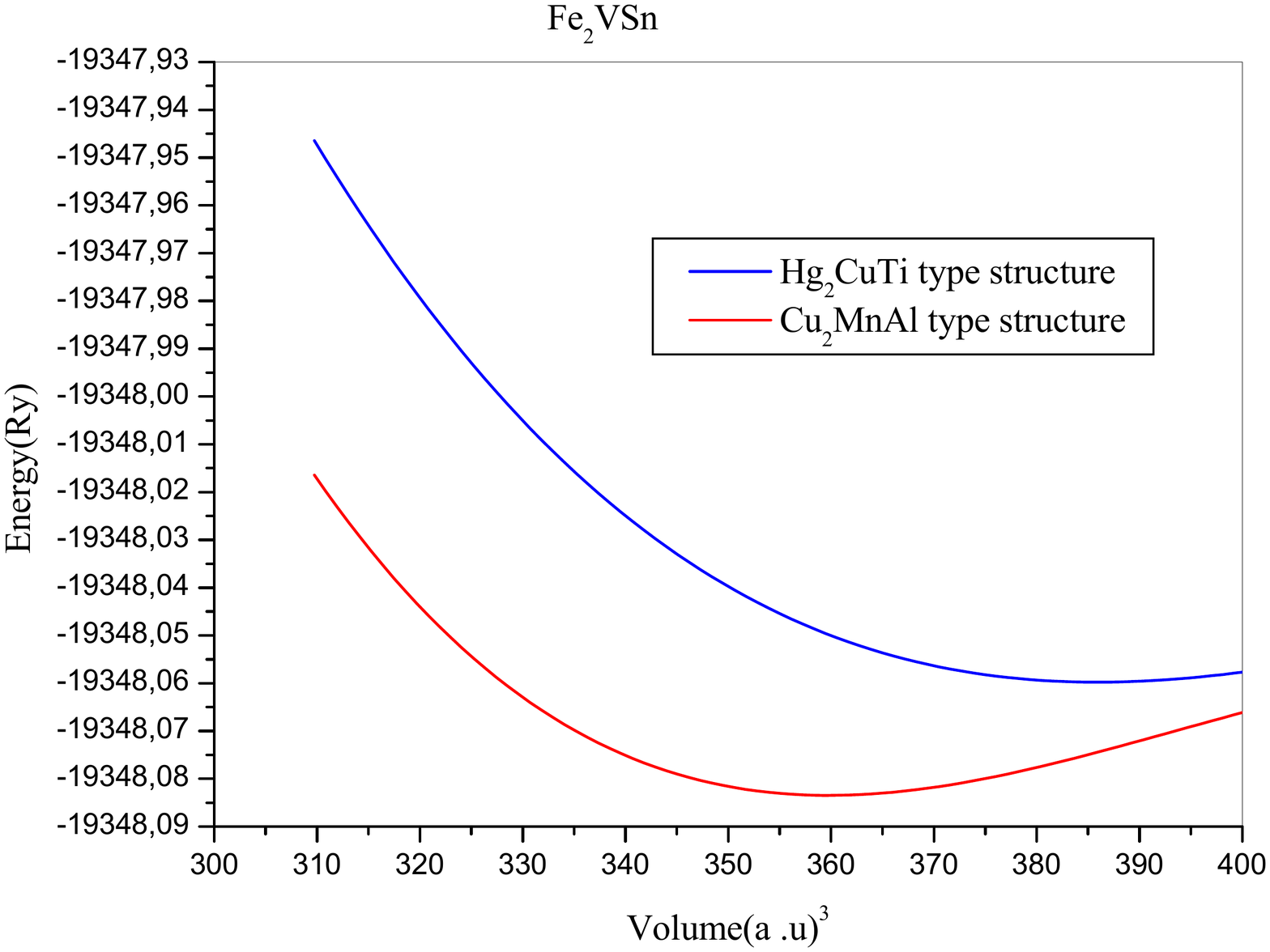} }}%
    \caption{(Colour online) The calculated total energy of Fe$_2$YSn (Y~=~Mn, Ti and V) alloys with both Cu$_2$MnAl and Hg$_2$CuTi types structures as a function of the lattice constants for magnetic states. }%
    \label{fig1}%
\end{figure}

There are three distinct families of Heusler compounds: the first one with the composition 1:1:1 and the second one with 2:1:1 stoichiometry and the third is 1:1:1:1. The compounds of the first family have the general formula XYZ and crystallize in a non-centro symmetric cubic structure; the second family of Heusler alloys has a formula X$_{2}$YZ with two types of structures the Hg$_{2}$CuTi and Cu$_{2}$MnAl. The two phases consist of four inter-penetrating fcc sub-lattices, which have four crystal sites, A(0,~0,~0), B(0.25,~0.25,~0.25), C(0.50,~0.50,~0.50) and D(0.75,~0.75,~0.75). 
%
%
For Hg$_{2}$CuTi type structure, the chain of atoms occupies the four sites of unit cell  X-X-Y-Z and for Cu$_{2}$MnAl the Y and the second X atom exchange sites.
In the Hg$_{2}$CuTi type structure, the X atoms entering sites A and B are denoted as X(1) and X(2), respectively~\cite{13}. The third family has a formula of XX'YZ and crystallize in the LiMgPdSn type crystal structure. For the Heusler alloys X$_{2}$YZ, the X and Y are both a transition metal, and Z is the main group element. In order to establish a stable structure and equilibrium structural parameters of Fe$_{2}$ZSn (Z~=~Mn, Ti and V) compounds, structural optimizations were performed on these alloys for both Cu$_{2}$MnAl and Hg$_{2}$CuTi type structures and their total energy-volume curves are shown in figure~\ref{fig1}. In X$_{2}$YZ Heusler alloys, if the Y atomic number is superior to that of X atom from the same period, an inverse Heusler structure with Hg$_{2}$CuTi type as the prototype is observed. It is seen from these E-V curves that the Cu$_{2}$MnAl type structure is more stable than the Hg$_{2}$CuTi phase for the Fe$_{2}$YSn with Y~=~Mn, Ti, V compounds at ambient conditions. The nuclear charge of X atom (Fe) is larger than Y atom (Y~=~Mn, Ti and V). Consequently, the Cu$_{2}$MnAl structure will be visibly observed as can be seen from figure~\ref{fig1}. The minimum of the curve is the calculated equilibrium lattice constant. The lattice constant $a$, bulk modulus $B$ and its pressure derivative $B'$ at zero pressure, for the structures Cu$_{2}$MnAl and Hg$_{2}$CuTi are calculated using Murnaghan equation of state~\cite{17}.

\begin{equation}
E(V)=E_{0}(V)+\frac{BV}{B'(B'-1)}\left[B\left(1-\frac{V_{0}}{V}\right)+\left(\frac{V_{0}}{V}\right)^{B'}-1\right].
\label{moneq}
\end{equation}
Here $E_0$ is the minimum energy at $T= 0$~K, $B$ is the bulk modulus, $B'$ is the bulk modulus derivative and $V_0$ is the equilibrium volume.
The results are listed in table~\ref{tbl-smp1}. The calculated lattice constants of Fe$_{2}$YSn with (Y~=~Mn, Ti and V) are in good agreement with the previously theoretically optimized lattice constants reported by other researchers.

 \begin{table}[htb]
\caption{The calculated equilibrium lattice constant $a$~(\AA), the bulk modulus $B$~(GPa), the minimum energy~(Ry) and the formation energy $E_f$~(Ry) of Fe$_2$YSn with (Y~=~Mn, Ti and V).}
\label{tbl-smp1}
\vspace{2ex}
\begin{center}

\renewcommand{\arraystretch}{0}
\begin{tabular}{|c|c||c|c|c|c|c||}
\hline

      & &$a$ (\AA) &$B$ (GPa)&$B'$&Energy (Ry)&$E_f$ (Ry)\strut\\
\hline
\rule{0pt}{2pt}&&&&&\\
\hline
\raisebox{-5.7ex}[0pt][0pt]{Fe$_2$TiSn}
   
 &  Hg$_2$CuTi&6.2173&124.0086&4.9226&$-19157.08477$&$-1.2578$\strut\\
   
\cline{2-7}
& \raisebox{-4.3ex}[0pt][0pt]{Cu$_2$MnAl}&6.0436&\raisebox{-4.3ex}[0pt][0pt]{191.5580}&\raisebox{-4.3ex}[0pt][0pt]{3.7601}&\raisebox{-4.3ex}[0pt][0pt]
{$-19157.15643$}&\raisebox{-4.3ex}[0pt][0pt]{$-1.2783$}\strut\\
& &6.03 \cite{18}&&&&\strut\\
& &6.04 \cite{19}&&&&\strut\\
& &6.04 \cite{20}&&&&\strut\\
  
\hline
\raisebox{-3.7ex}[0pt][0pt]{Fe$_2$VSn}
      
 &  \raisebox{-1.2ex}[0pt][0pt]{Hg$_2$CuTi}&6.1239&\raisebox{-1.2ex}[0pt][0pt]{133.9734}&\raisebox{-1.2ex}[0pt][0pt]{5.0141}&\raisebox{-1.2ex}[0pt][0pt]{$-19348.05986$}&\raisebox{-1.2ex}[0pt][0pt]{$-1.6761$}\strut\\
& &6.06 \cite{21}&&&&\strut\\

\cline{2-7}
     
& \raisebox{-1.2ex}[0pt][0pt]{Cu$_2$MnAl}&5.9824&\raisebox{-1.2ex}[0pt][0pt]{142.7406}&\raisebox{-1.2ex}[0pt][0pt]{9.9780}&\raisebox{-1.2ex}[0pt][0pt]{$-19348.08220$}&\raisebox{-1.2ex}[0pt][0pt]{$-1.6985$}\strut\\
& &5.99 \cite{21}&&&&\strut\\

\hline
\raisebox{-3.7ex}[0pt][0pt]{Fe$_2$MnSn}

 &  Hg$_2$CuTi&6.0843&110.4683&6.0807&$-19766.69924$&$-1.3350$\strut\\
 
\cline{2-7}
       
& \raisebox{-2.7ex}[0pt][0pt]{Cu$_2$MnAl}&5.9584&\raisebox{-2.7ex}[0pt][0pt]{178.1479}&\raisebox{-2.7ex}[0pt][0pt]{5.2306}&\raisebox{-2.7ex}[0pt][0pt]{$-19766.69677$}&\raisebox{-2.7ex}[0pt][0pt]{$-1.3425$}\strut\\
& &5.70 \cite{22}&&&&\strut\\
& &6.01 \cite{23}&&&&\strut\\

\hline

\end{tabular}
\renewcommand{\arraystretch}{1}
\end{center}
\end{table}

We study the phase stability of Fe$_2$YSn with (Y~=~Mn, Ti, V) based on the formation energy ($\Delta E_f$). This can help to envisage whether these alloys can be prepared experimentally. Here, the formation energy ($\Delta E_f$) is calculated by comparing the total energies of the Fe$_2$YSn (Y~=~Mn, Ti, and V) Heusler alloys with the sum of the total energies of the constituting elements.
The formation energy of the Fe$_2$YSn (Y~=~Mn, Ti, and V) materials is computed following the expression given below:

\begin{equation}
\Delta E_f=E^{\text{total}}_{\text{Fe}_2\text{YSn}}-\left[E^{\text{bulk}}_{\text{Fe}}+E^{\text{bulk}}_{\text{Y}}+E^{\text{bulk}}_{\text{Sn}}\right].
\label{moneq}
\end{equation}
\begin{figure}[!t]	
	\centering
	\includegraphics[width=8.7cm]{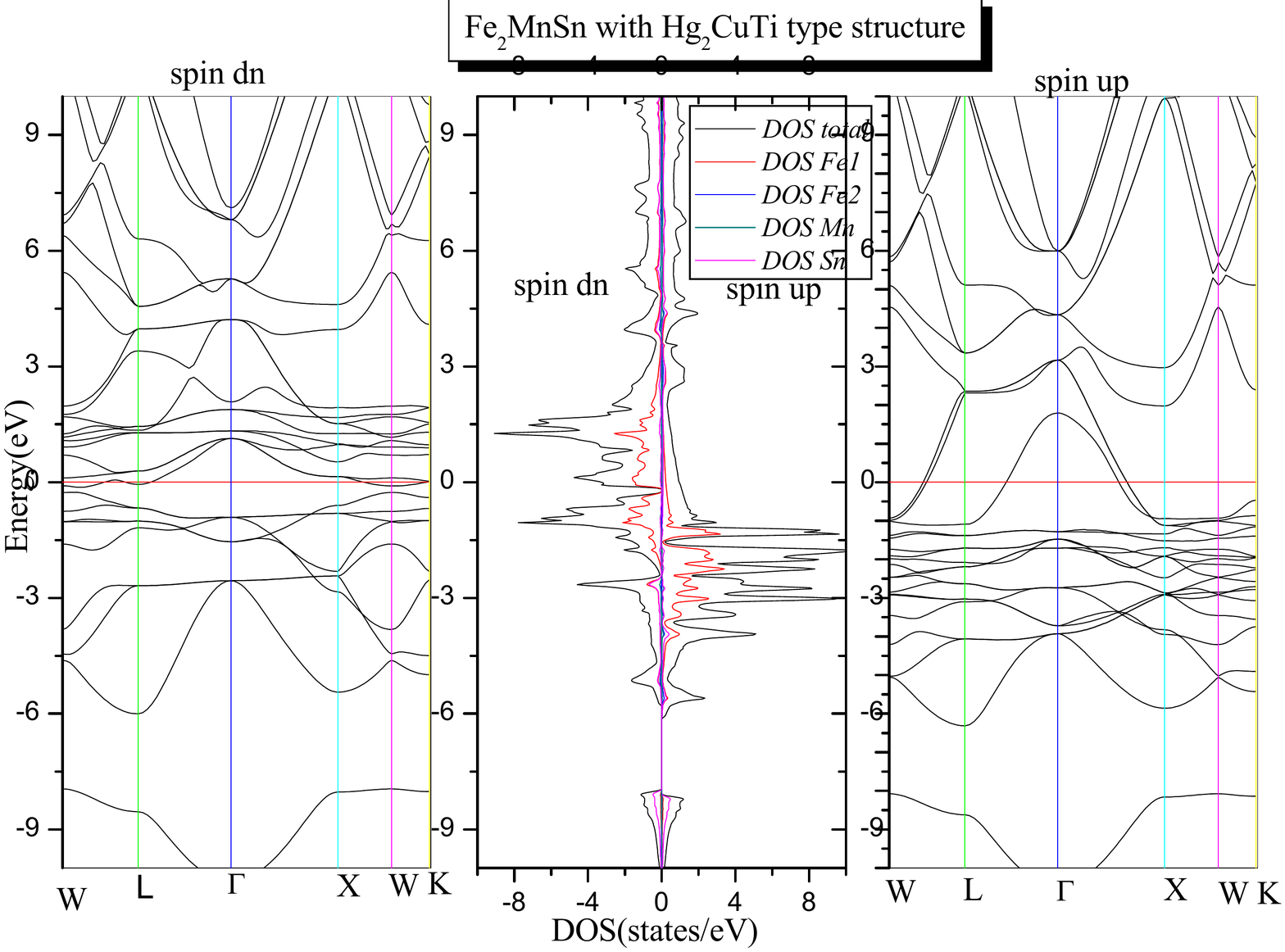}
	\includegraphics[width=8.7cm]{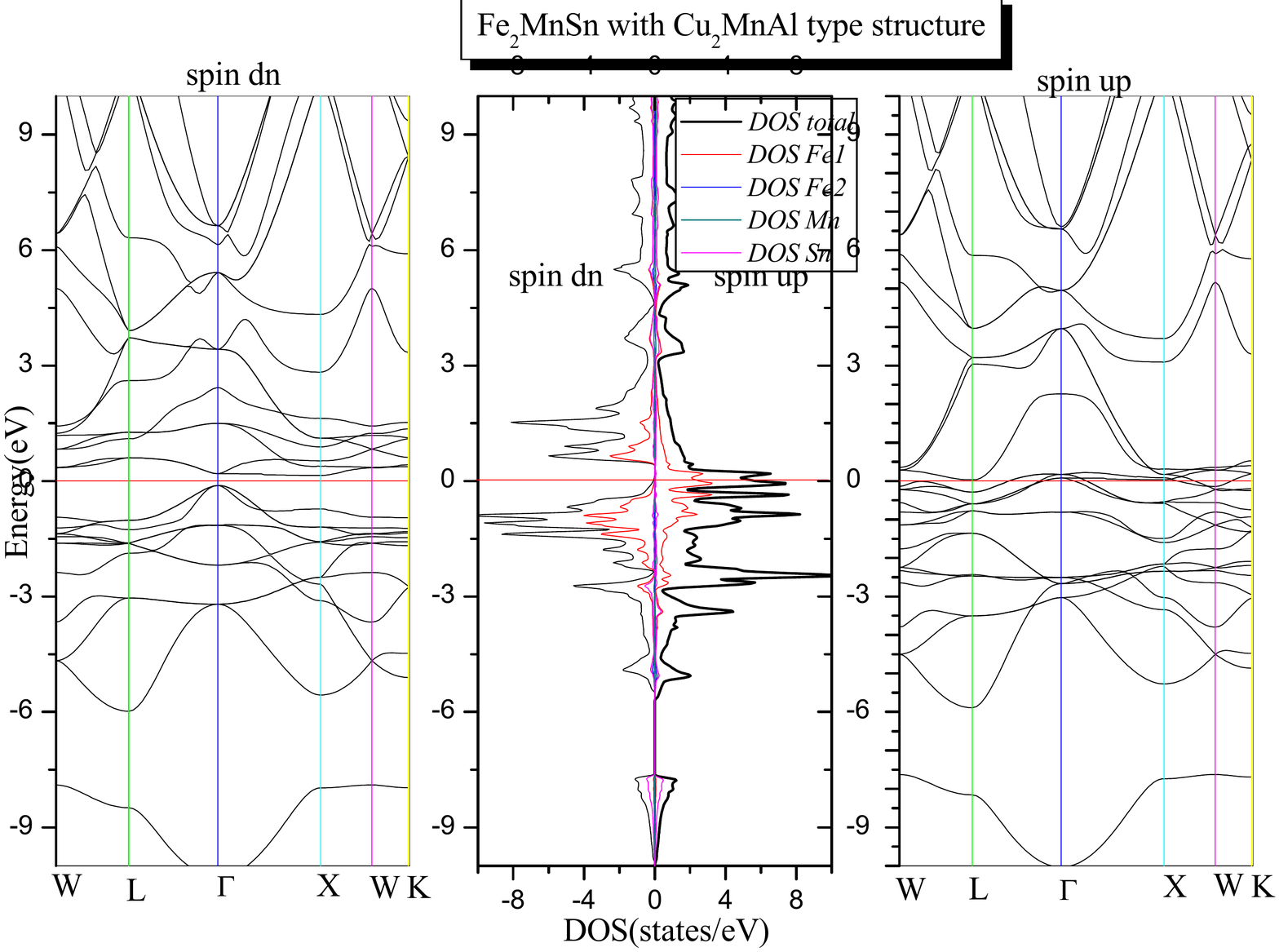}
	\caption{(Colour online) The calculated spin-polarized total and partial density of states and band structure of Fe$_2$MnSn with both Cu$_2$MnAl and Hg$_2$CuTi types structures.}  
	\label{fig2}
\end{figure}
Here $E^{\text{total}}_{\text{Fe}_2\text{YSn}}$ is the energy of the Hg$_2$CuTi type and Cu$_2$MnAl type structure under their equilibrium lattice constant for the Fe$_2$YSn with (Y~=~Mn, Ti and V) full Heusler alloys, and $E^{\text{bulk}}_{\text{Fe}}$, $E^{\text{bulk}}_{\text{Y}}$, $E^{\text{bulk}}_{\text{Sn}}$ represent the total energy per atom for Fe, Y~(Y~=~Mn, Ti, V), and Sn elements in the bulk form, respectively. The negative values of the formation energy specify that Fe$_2$YSn with (Y~=~Mn, Ti and V) full Heusler alloys are chemically stable, and these materials can be synthesized experimentally. The computed results of the Cu$_2$MnAl type structures are found more negative than those of the Hg$_2$CuTi types structures, proving that Cu$_2$MnAl type structures are more stable compared to the Hg$_2$CuTi type structures.
In half-metallic Heusler compounds, the gap takes place in one spin state, whereas in the other spin state, EF cuts through the bands~\cite{24}. The $d$-band is principally responsible for the position of Fermi level lying in it. The responsibility of transition metals 3$d$-states is very important in the description of spin polarized electronic band structures and densities of sates calculations~\cite{24}.
At the equilibrium lattice constants, we have studied the electronic band structure calculations for all three compounds and have extracted the density of states (DOS) per f.u., which is presented in figure~\ref{fig2}, figure~\ref{fig3} and figure~\ref{fig4}.
The electronic band structure shows the bonding and character of the electron bands. DFT is a standard tool for calculating the band structure for materials in order to determine different properties of solids~\cite{25,26}. The responsibility of transition metals 3$d$-states is very essential in the description of spin polarized electronic band structures and densities of sates calculations~\cite{24}.
The electron spin polarization (SP) at EF of a material is defined as follows~\cite{23}

\begin{equation}
SP=\frac{\rho_\uparrow (E_f)- \rho_\downarrow (E_f) }{\rho_\uparrow (E_f) + \rho_\downarrow(E_f)}.
\label{moneq}
\end{equation}
Here $\rho_\uparrow (E_f)$, $ \rho_\downarrow (E_f)$ are the majority and minority densities of states at $E_f$. When the value of the electron spin polarization (SP) is 100\%, alloys are supposed to be true half-metallic, and this is realized when any one of the DOS from the majority and minority spins is equal to zero and the other one is not equal to zero at EF~\cite{23}.
%
%
Figure~\ref{fig2} presents the total density of states and band structure of Fe$_2$MnSn for both Cu$_2$MnAl and Hg$_2$CuTi  type structures.
From this figure, one can observe that the alloy exhibits half-metallic behavior with Cu$_2$MnAl type structure, described by an overlap between the bottom of the conduction band and the top of the valence band in spin-up. For spin-down, we can see a gap between the maximum of the valance band and the minimum of the conduction band, this gap being indirect between $\Gamma$ and X points. For Hg$_2$CuTi type structure, both the majority and minority spin bands have metallic intersections at the Fermi level.
We can see from figure~\ref{fig3} that the total DOS in spin-up and spin-down spin channels for  Fe$_2$TiSn, is symmetrical in the majority and minority spin directions. Therefore, the non-magnetic character of these alloys can be estimated. In both spin directions, the energy gap is open around the Fermi level EF and divides the total DOS into bonding and anti-bonding parts. The formation of this gap is related to the hybridization of Fe and Ti $d$ electrons. There is no contribution from titanium (Ti) and tin (Sn) to the density of states at the Fermi level. We presented the energy bands of Fe$_2$TiSn in figure~\ref{fig3}. It can be seen that the energy gap in Fe$_2$TiSn is an indirect gap and the Fermi level locates just above the top of the valence band at the $\Gamma$ point.
The electronic band structures and the DOS of the Fe$_2$VSn compounds for both the Hg$_2$CuTi and Cu$_2$MnAl type structures are given in figure~\ref{fig4}, the density of states of spin-up and spin-down occurs at Fermi level; as a result, Fe$_2$VSn is of a metallic character. Both the conduction and valence bands cross the Fermi level, thus diminishing the gap at EF. The metallic nature in this compound is principally due to the interaction between Sn-$p$ and transition metal (TM)-3$d$ states.
The Fe-3$d$ and Y-3$d$ (Y~=~Mn, Ti, V) states are mainly occupied around the Fermi level with a maximum contribution towards the total DOS and, as a result, the corresponding bonding-anti-bonding states control the energy gap formation~\cite{27}. At the same time, the Sn atomic states are less active around the Fermi level in these materials. Thus, the observed band gap in these alloys is due to the typical $d-d$ hybridization between the valence states of Fe and Y atoms (Y~=~Mn, Ti and V).
Skaftouros et al.~\cite{28} have presented fascinating arguments regarding possible hybridizations between $d$-orbitals of transition metals in the case of the X$_2$YZ Inverse Heusler compounds, e.g., Sc-based Heusler compounds. According to their report, the same symmetry of the X~\cite{1} and the Y atoms causes their $d$-orbitals to hybridize together creating five bonding $d$ ($2 \times e_g$ and $3 \times t_{2g}$) and five non-bonding ($2 \times e_u$ and $3 \times t_u$) states. Then, the five X(1)-Y bonding $d$ states hybridize with the $d$-orbital of the X(2) atoms and create bonding and anti-bonding states again (3).

\begin{figure}[!t]	
	\centering
	\includegraphics[width=9cm]{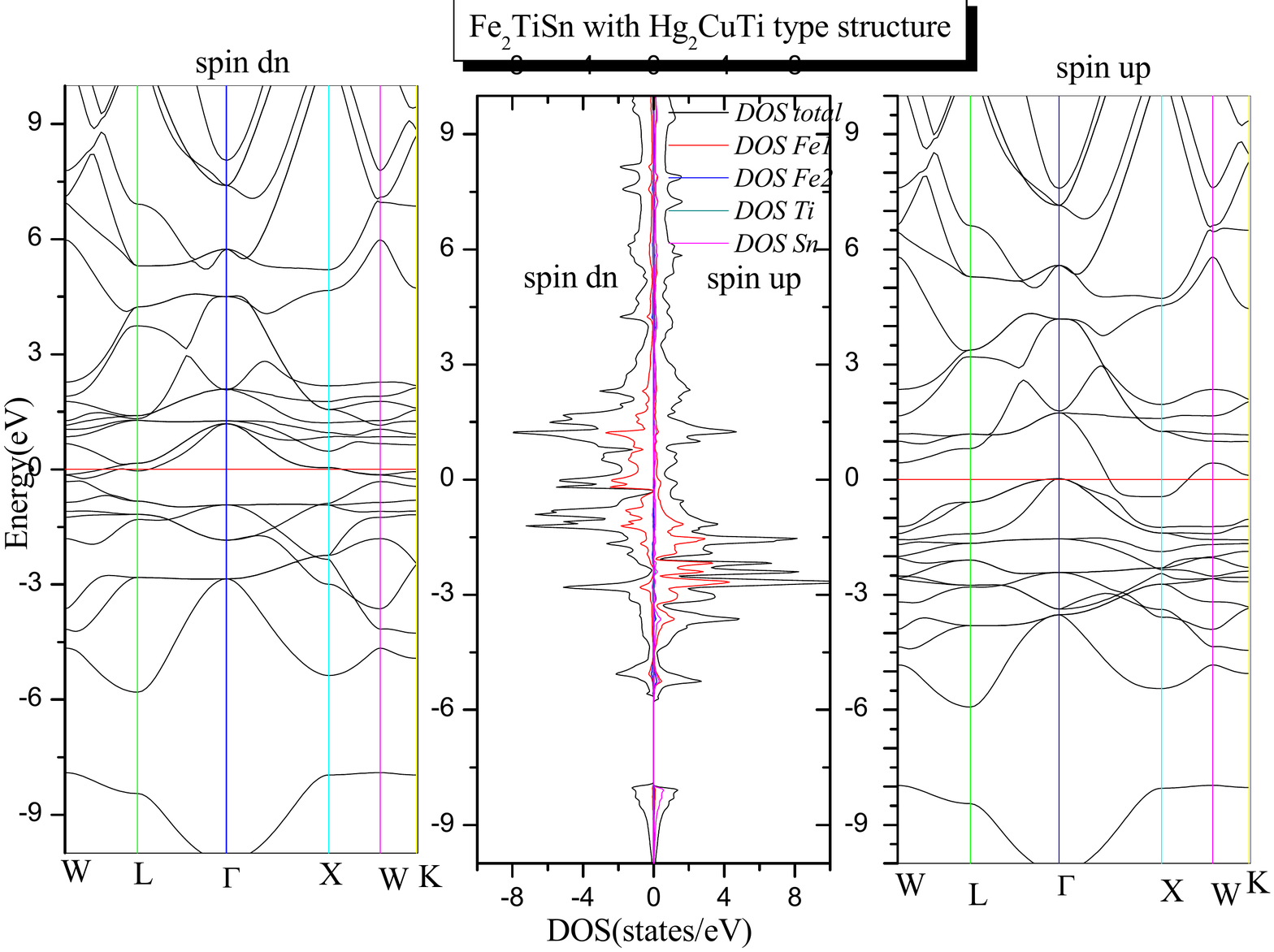}
	\includegraphics[width=9cm]{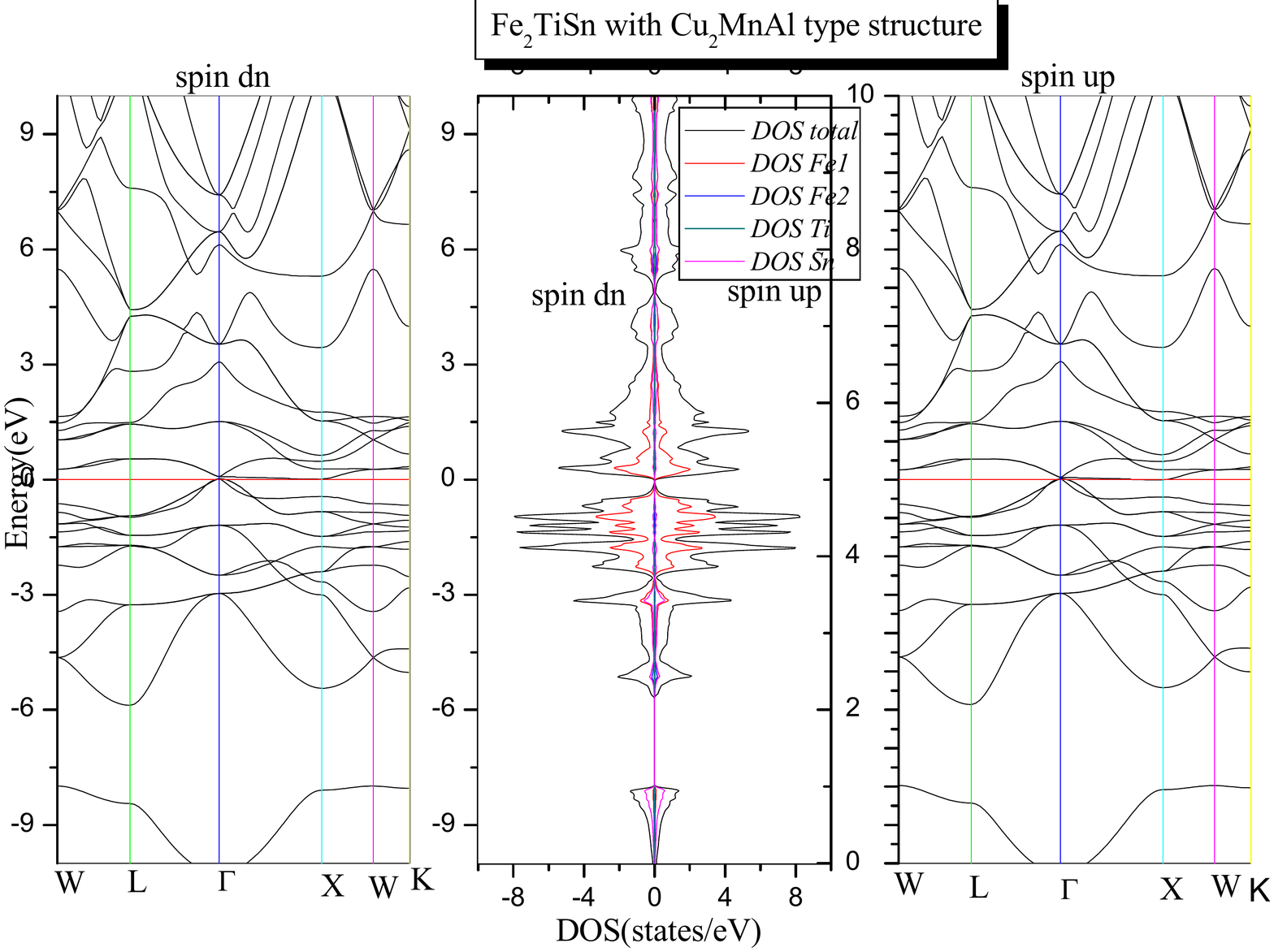}
	\caption{(Colour online) The calculated spin-polarized total and partial density of states and band structure of Fe$_2$TiSn with both Cu$_2$MnAl and Hg$_2$CuTi types structures.}  
	\label{fig3}
\end{figure}

 
\begin{figure}[!t]	
	\centering
	\includegraphics[width=8.7cm]{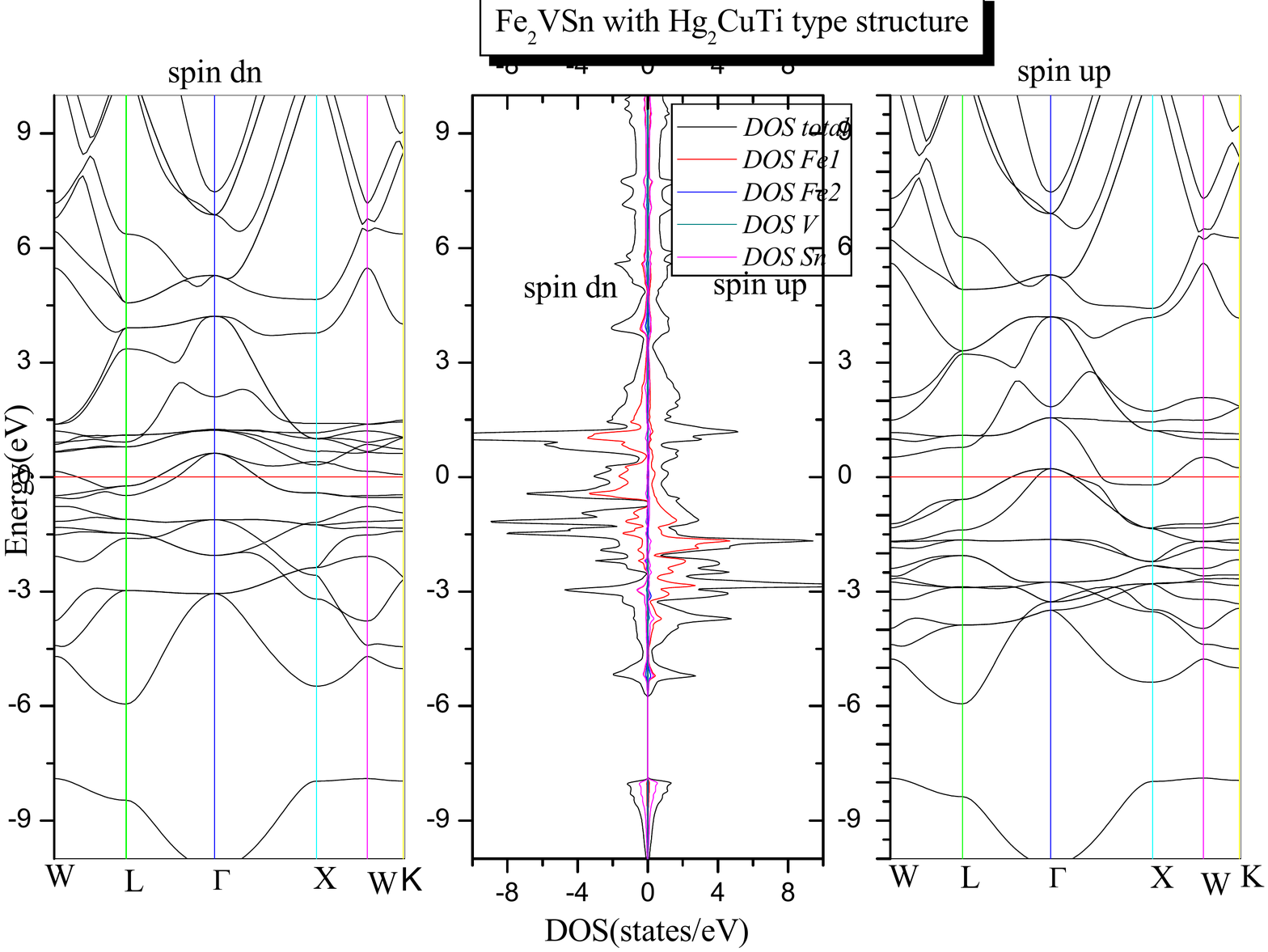}
	\includegraphics[width=8.7cm]{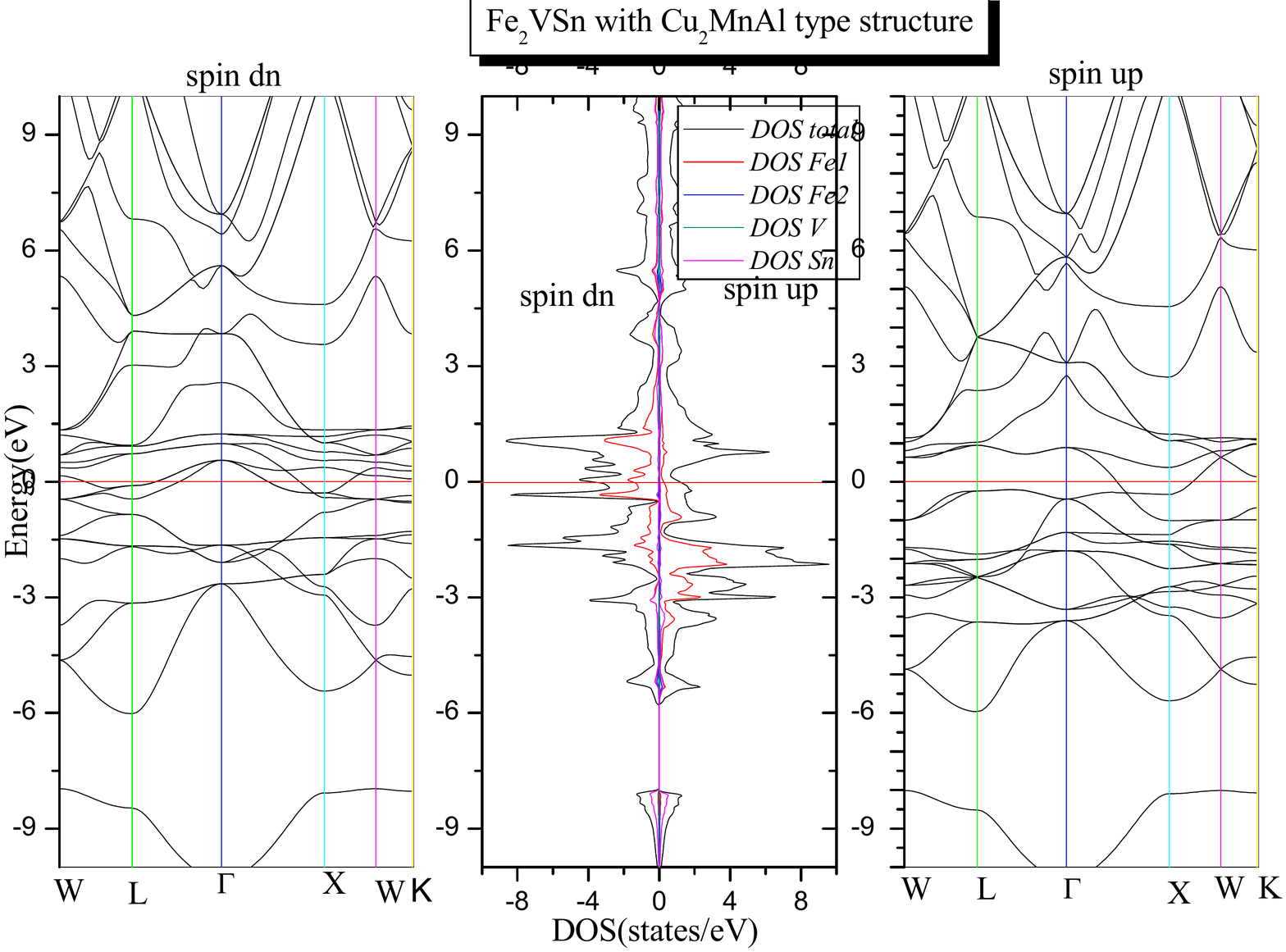}
	\caption{(Colour online) The calculated spin-polarized total and partial density of states and band structure of Fe$_2$VSn with both Cu$_2$MnAl and Hg$_2$CuTi types structures.}  
	\label{fig4}
\end{figure}


 \begin{table}[!t]
\caption{The calculated magnetic moments values ($\mu_\text{B}$) of the of  Fe$_2$YSn (Y~=~Mn, Ti and V) Heusler compounds.}
\label{tbl-smp2}
\vspace{2ex}
\begin{center}

\renewcommand{\arraystretch}{0}
\begin{tabular}{|c|c||c|c|c|c|c|c||}
\hline
    
 & &$M_{\text{Fe}_1}$&$M_{\text{Fe}_2}$&$M_Z$&$M_{\text{Sn}}$&$M_{\text{interstitial}}$&$M_ {\text{total}}$ ($M_t$)\strut\\

\hline
\rule{0pt}{2pt}&&&&&\\
\hline
\raisebox{-1.7ex}[0pt][0pt]{Fe$_2$MnSn}
   
 &  Hg$_2$CuTi&2.47398&2.66903&3.29914&-0.06182&-0.05092&8.32940\strut\\
    
\cline{2-8}

& Cu$_2$MnAl&$-0.20534$&$-0.16421$&3.37041&$-0.00477$&0.01605&3.00014\strut\\
  
\hline
\raisebox{-2.7ex}[0pt][0pt]{Fe$_2$TiSn}
    
 &  Hg$_2$CuTi&2.34182&2.49429&$-0.46420$&$-0.00535$&$-0.21745$&4.14910\strut\\

\cline{2-8}
     
& \raisebox{-1.1ex}[0pt][0pt]{Cu$_2$MnAl}&0.00092&\raisebox{-1.1ex}[0pt][0pt]{0.00053}&0.00023&0.00001&\raisebox{-1.1ex}[0pt][0pt]{0.00014}&0.00000\strut\\
&&0.00 \cite{29}&&0.00 \cite{29}&0.00 \cite{29}&&0.00 \cite{29}\strut\\

\hline
\raisebox{-1.7ex}[0pt][0pt]{Fe$_2$VSn}

 &  Hg$_2$CuTi&2.07363&2.73598&$-1.63655$&$-0.00579$&$-0.40861$&2.75866\strut\\
 
\cline{2-8}
       
& Cu$_2$MnAl&2.19361&2.19754&$-1.06404$&$-0.01341$&$-0.28226$&3.23143\strut\\

\hline

\end{tabular}
\renewcommand{\arraystretch}{1}
\end{center}
\end{table}

We present the total magnetic moment, the local magnetic moments on Fe, Y~(Y~=~Mn, Ti, and V), Sn atoms and interstitial moments which are given per unit cell.
The calculated local and total magnetic moments in interstitial region for a Heusler compound Fe$_2$YSn with (Y~=~Mn, Ti and V) are presented in table~\ref{tbl-smp2}. It must be noted that the total magnetic moment is very sensitive to both types of structures.
For Fe$_2$YSn with (Y~=~Mn, Ti and V) with Hg$_2$CuTi type structure, the magnetic moment is mostly located on the Fe atoms. Indeed, a great part of the total magnetic moment results from this atom. Therefore, the Sn atom has a minor magnetic moment, which does not give a lot to the total magnetic moment.
The found values of the magnetic moment are dependent on the Slater-Pauling curve (SPC)~\cite{30} for full Heusler alloys, in which the magnetic moment per unit cell in multiples of Bohr magnetons ($\mu_\text{B}$) can be calculated as follows:

\begin{equation}
M_{\text{tot}}=N-24.
\label{moneq}
\end{equation}
Here $M_\text{tot}$ represents the total magnetic moment and N represents the total valence electrons in the unit cell. $N$ is equal to 27 for  Fe$_2$MnSn [$(8\times 2)+7+4=27$], it is equal to 24 for Fe$_2$TiSn and [$(8\times 2)+4+4=24$]. The value of calculated magnetic moment have an integer value of 0$\mu_\text{B}$ and 3$\mu_\text{B}$ for Fe$_2$TiSn and Fe$_2$MnSn respectively with Cu$_2$MnAl type structure, which matches well with moments predicted from Slater-Pauling rule and evidences that these have potential to be half metallic.
The magnetic moment of Fe$_2$TiSn is zero, the interaction between $3d$ electron of Fe and $3d$ electrons of Ti are in opposite directions and cancel the over-all moment.
The Heusler alloy Fe$_2$TiSn has 24 valence electrons. These electrons occupy the majority and minority spin bands equally (12 up and 12 down), which results in a nonmagnetic semiconductor-like band structure~\cite{29}. These results agree well with preceding studies in Fe$_2$TiSn~\cite{29}.
The total magnetic moment of Fe$_2$VSn is equal to 2.75866$\mu_\text{B}$ and 3.23143$\mu_\text{B}$ for Hg$_2$CuTi and Cu$_2$MnAl type structure very far from the integer value witch confirm that this alloys have not half metallic behavior.

\section{Conclusion}\label{s4}
We have performed ab-initio calculations to investigate the structural, electronic, magnetic properties of Fe$_2$YSn (Y~=~Mn, Ti and V) Heusler alloys with both Cu$_2$MnAl and Hg$_2$CuTi type structure. The negative formation energy is shown, as an evidence of the thermodynamic stability of Fe$_2$YSn (Y~=~Mn, Ti, and V) alloy. The Cu$_2$MnAl type structure is energetically more stable than the Hg$_2$CuTi type structure. Our calculations indicate that the 6.08~\AA, 5.98~\AA~and 5.95~\AA~are the equilibrium lattice constant of Fe$_2$TiSn,  Fe$_2$VSn and Fe$_2$MnSn with Cu$_2$MnAl, respectively. Furthermore, Fe$_2$MnSn and  Fe$_2$TiSn in the ground state is considered to be a true half-metallic based on the calculations of the band structure and density of states. It  is also predicted that Fe$_2$MnSn and  Fe$_2$TiSn compounds are half-metallic with 100\% spin
 polarization with an integer magnetic moments making these compounds a good candidates for spintronic devices applications.

\section{Acknowledgement}

This work has been supported by DGRST-ALGERIA.


\ukrainianpart

\title[Першопринципне моделювання напівметалічного феромагнетизму сплавів Гейслера...,]
{Першопринципне моделювання напівметалічного феромагнетизму сплавів Гейслера Fe$_2$YSn (Y=Mn, Ti та V)}
\author[М. Саях \textsl{та ін.}]{M. Саях\refaddr{label1}, С. Зеффане \refaddr{label1}, М. Мохтарі\refaddr{label1,label2}, Ф. Дахмане\refaddr{label1,label3}, Л. Зекрі\refaddr{label2}, Р. Хената\refaddr{label3}, Н.~Зекрі\refaddr{label2}}
\addresses{
	\addr{label1} Інститут науки і техніки, університетський центр Тіссемсілта, 38000 Тіссемсілт, Алжир
	\addr{label2} Університет науки і техніки ім. Мухаммеда Будіафа в Орані, Ель Нуар, 31000 Оран, Алжир
	\addr{label3} Лабораторія квантової фізики та математичного моделювання, Технологічний факультет Університету Маскара, 29000 Маскара, Алжир }

\makeukrtitle 
\begin{abstract}
У роботі для дослідження структурних, електронних та магнітних властивостей Fe$_{2}$YSn з (Y~=~Mn, Ti та V) використовується першопринципне моделювання на основі функціоналу густини у рамках узагальненого градієнтного наближення. Cтруктури типу Cu$_{2}$MnAl є більш енергетично стійкими, ніж Hg$_{2}$CuTi. Від'ємна енергія утворення є свідченням термодинамічної стійкості сплаву. Для рівноважного значення сталої гратки розрахований повний спіновий момент складає 3$\mu_\text{B}$ та 0$\mu_\text{B}$ для сплавів Fe$_{2}$MnSn та Fe$_{2}$TiSn, відповідно, що узгоджується з правилом Слейтера-Паулінга $M_t= Z_t-24$. Вивчення електронних та магнітних властивостей підтверджує, що повні гейслерівські сплави Fe$_{2}$MnSn і Fe$_{2}$TiSn  є  типовими напівметалічними феромагнітними матеріалами.
\keywords cплави Гейслера, електронна структура, першопринципне моделювання, напівметалічність
\end{abstract}

\end{document}